The ARF-AID system: Methods that preserve endogenous protein levels and facilitate rapidly inducible protein degradation

Kizhakke Mattada Sathyan[1*], Thomas G. Scott[1], and Michael J. Guertin[1,2,3]

[1]Biochemistry and Molecular Genetics Department, [2]Center for Public Health Genomics, [3]Cancer Center, University of Virginia, Charlottesville, Virginia 22908, USA

KMS: sk8fz@virginia.edu
TGS: ts2hx@virginia.edu
MJG: guertin@virginia.edu
*Correspondence

**ABSTRACT:**

The ARF-AID (Auxin Response Factor-Auxin Inducible Degron) system is a re-engineered auxin-inducible protein degradation system. Inducible degron systems are widely used to specifically and rapidly deplete proteins of interest in cell lines and organisms. An advantage of inducible degradation is that the biological system under study remains intact and functional until perturbation. This feature necessitates that the endogenous levels of the protein are maintained. However, endogenous tagging of genes with AID can result in chronic, auxin-independent proteasome-mediated degradation. The additional expression of the ARF-PB1 domain in the re-engineered ARF-AID system prevents chronic degradation of AID-tagged proteins while preserving rapid degradation of tagged proteins. Here we describe the protocol for engineering human cell lines to implement the ARF-AID system for specific and inducible protein degradation. These methods are adaptable and can be extended from cell lines to organisms.

Basic Protocol 1: (Generation of eGFP-ARF-P2A-TIR1 or ARF-HA-P2A-TIR1 progenitor cells)

Basic Protocol 2: (Tagging a gene of interest with AID)

Basic Protocol 2: (Testing auxin-mediated degradation of the AID-tagged protein)

**INTRODUCTION:**

A diversity of molecular tools that disrupt genes are commonly used to gain mechanistic insight into protein function. Many of the methods available today disrupt gene function by genetic knock-out or RNA degradation. These methods can be universally applied to study most genes and allow us to understand the cumulative effect of gene dysregulation. The major drawbacks of these systems are that the kinetics of protein depletion are slow and chronic, ranging from days to the lifetime of an organism, and often irreversible. This is problematic when studying the mechanistic function of a protein, which is most directly assessed by observing the immediate cellular response to dysregulation. Moreover, chronic gene disruption is not possible for essential genes. Small molecule inhibitors and temperature-sensitive mutations are acute, rapid, and reversible, but unique strategies are needed to target each protein of interest. Inducible degron systems are rapid, reversible, and can be universally applied to any protein.

One strategy to directly manipulate protein stability is to induce interaction with a ubiquitin ligase complex, which will lead to polyubiquitination and proteasomal degradation of the protein (Sakamoto et al., 2001; Schneekloth et al., 2008; Schapira et al., 2019). PROTACs (Proteolysis Targeting Chimeras) are heterobifunctional molecules that promote proximity-mediated polyubiquitination. PROTACs are composed of a moiety that binds to an E3 ubiquitin ligase, such as von Hippel-Lindau (VHL) or cereblon (CRBN), and a small molecule that directly interacts with the protein of interest (Bondeson et al., 2015; Lu et al., 2015; Winter et al., 2015; Schapira et al., 2019). This strategy requires a chemical probe for the protein of interest as starting material, and developing PROTACs for each target is time-consuming and requires medicinal chemistry expertise.

The dTAG system provides a more universal system to specifically target proteins of interest for rapid and inducible ubiquitin-mediated degradation. In the dTAG system the protein of interest is fused to a mutant human $FKBP^{F36V}$, and a single bifunctional molecule promotes proteasome targeting (Nabet et al., 2018). This system is simple and requires only one genetic manipulation in order to tag the protein with $FKBP^{F36V}$. However, the degradation rate using this system varies depending on the cell type (Li et al., 2019). Additionally, the amount of the dTAG-13 molecule must be titered based on protein levels in order to avoid saturating each end of the molecule independently and not providing a link between the target and the ubiquitin ligase (Nabet et al., 2018; Li et al., 2019).

Other chemical genetics approaches to targeted protein degradation utilize the exogenous expression of plant-specific E3 ubiquitin ligase adaptor proteins in animals and cell lines. The auxin inducible degron (AID) system was the first heterologous system developed (Nishimura et al., 2009). In this system an auxin molecule interacts with the TIR1 protein, which acts as a ubiquitin ligase adapter. This auxin-induced interaction of AID with the SCF-TIR1 E3 ubiquitin ligase complex causes ubiquitination and degradation of the AID-tagged protein (Nishimura et al., 2009). This was followed by the development of the Jasmonate inducible degron (JID) system. Here, in the presence of jasmonate-isoleucine, proteins tagged with the JAZ degron interact with the F-box containing COI1 and are subsequently degraded (Brosh et al., 2016). Recently, another auxin-sensing F-box protein *A. thaliana* AFB2 (AtAFB2) has been developed as a promising new degron system (Li et al., 2019). Of the direct protein degradation technologies, the auxin-inducible degron system is the most robust and most widely used system (Lambrus et al., 2018).

Endogenously tagging genes with AID can result in unwanted chronic degradation in the absence of auxin. Supplementing the AID system with an additional component of the plant's native auxin signaling machinery preserves near-endogenous expression levels of the target protein (Sathyan et al., 2019). The canonical AID system has two components: 1) transport inhibitor response 1 (TIR1), and 2) auxin/indole-3-acetic acid (Aux/IAA or AID) proteins (Nishimura et al., 2009). However, in plants, there is another critical component in the auxin signal transduction system: auxin response transcription factors (ARF). In the absence of auxin, ARF binds to the AID protein and protects it from TIR1-mediated ubiquitination. Upon sensing auxin, TIR1 binds to and ubiquitinates the AID protein, which dissociates from ARF (Dharmasiri et al., 2003, 2005; Gray et al., 2001). Introduction of the ARF-PB1 domain in the re-engineered auxin-inducible degron system (ARF-AID) rescued chronic auxin-independent degradation of AID-tagged proteins and increased the rate of auxin-induced AID-tagged protein degradation (Sathyan et al., 2019). A caveat of any auxin system is that the auxins are aromatic hydrocarbon molecules and indole-3-acetic acid (auxin) can cause changes in expression of aryl hydrocarbon receptor genes (Sathyan et al., 2019). We look forward to mixing and matching the components of newly developed AID systems in order to further refine these tools.

Simultaneous expression of ARF and TIR1 driven by a robust common promoter ensures high expression of these proteins compared to most cellular proteins (Figure 1 and Figure 2). We generated two multi-cistronic plasmids that express both ARF and TIR1 driven by a CMV promoter (Sathyan et al., 2019). A P2A ribosome skipping site separates these two polypeptides during translation (Figure 2A). In our original work, we used a CMV-driven eGFP-ARF to rescue the chronic degradation of AID-tagged proteins (Sathyan et al., 2019). In that context, we found that the rescued AID-tagged proteins degraded faster when treated with auxin (Sathyan et al., 2019). However, when we first expressed ARF-HA-P2A-TIR1 and then tagged proteins with AID in order to preserve protein levels, the degradation rate was slower compared to eGFP-ARF rescued degradation (Figure 3A,C&E). Therefore, we modified the ARF-HA construct to express eGFP-ARF (Addgene #129668). Both eGFP-ARF and ARF-HA preserve endogenous expression of the tagged proteins and promote comparably rapid degradation kinetics (Figure 3).

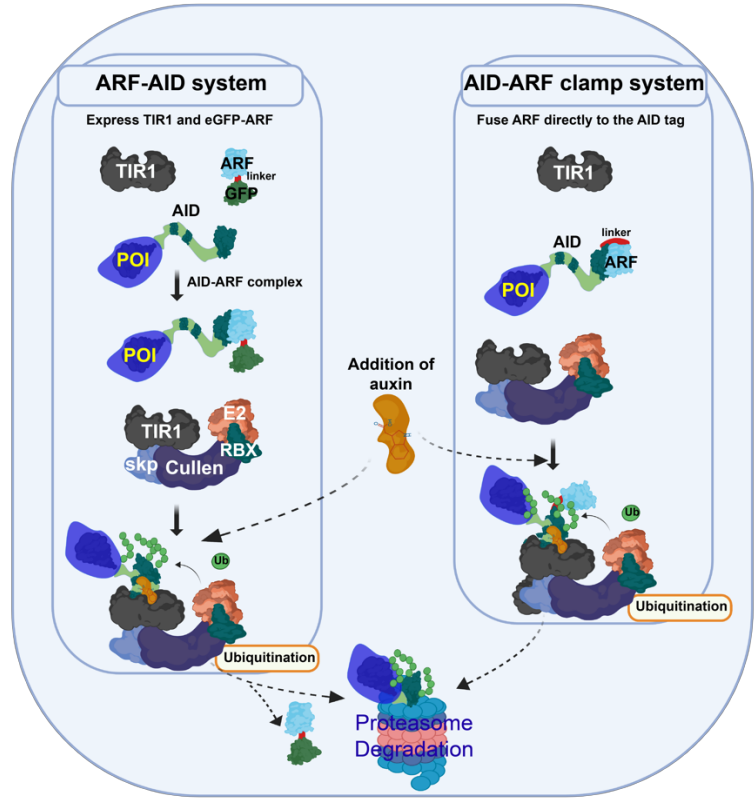

Figure 1. **An overview of the components and their roles in engineered ARF-AID systems**. The ARF-AID system (left) differs from traditional AID systems due to the presence of the ARF-PB1 domain, which binds to the AID-tag and prevents auxin-independent AID degradation. The AID-ARF clamp system (right) fuses the ARF-PB1 domain to the AID-tag, which also protects AID from auxin-independent degradation. An advantage of the clamp system is that previously generated TIR1-expressing cells and animals can be used as the progenitors for protein tagging. In both systems, auxin facilitates interaction between AID and TIR1 and rapid proteasome-mediated protein degradation.

ARF prevents chronic degradation of the AID-tagged proteins by direct interaction with AID (Sathyan et al., 2019). Therefore, it is important that ARF and the AID-tagged protein are localized to the same subcellular region. The ARF-TIR1 plasmids described here are designed for protein localized in the nucleus, so a nuclear localization signal (NLS) sequence is fused to the ARF-PB1 domain (Sathyan et al., 2019). In order to adopt the system to degrade cytoplasmic proteins, one should replace the NLS with a nuclear export signal.

Stable expression of ARF and TIR1 ensures efficient auxin-inducible degradation of the AID-tagged proteins. Integrating these genes at a safe harbor genetic locus allows ARF-PB1 and TIR1 to be stably expressed and resistant to epigenetic silencing (Figure 1 and Figure 2). Virus-mediated integration of the constructs at random genetic loci may lead to variable and unstable expression of ARF and TIR1. We incorporate ARF and TIR1 (eGFP-ARF-P2A-TIR1 or ARF-HA-P2A-TIR1) into the human AAVS1 locus.

Redesigning the eGFP-ARF-P2A-TIR1 plasmid with ROSA26-specific homology arms and using a mouse ROSA26 specific sgRNA (Chu et al., 2016) will allow for integration into mouse cells. For cells from other organisms, design a new sgRNA to the safe harbor locus and eGFP-ARF-P2A-TIR1 or ARF-HA-P2A-TIR1 with corresponding right and left homology arms. The plasmids have approximately 800 nucleotide long homology arms. Shorter homology arms, as few as 30 nucleotides, also permit efficient homology directed repair and have the advantage of increased transfection efficiency (Paix et al., 2017). We recommend generating a clonal progenitor cell line that expresses ARF and TIR1, then using this progenitor to tag proteins of interest (Figure 2A). We provide human-specific codon optimized constructs, but codon optimization is recommended for expression in other organisms.

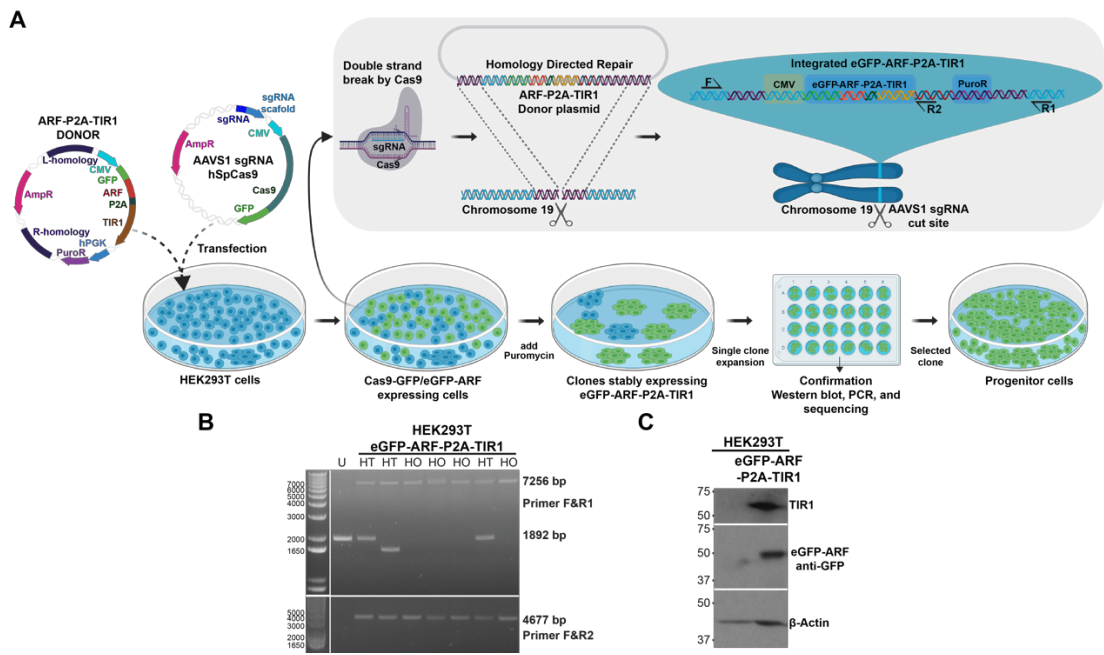

Figure 2. **Generating a stable progenitor cell line that expresses the components of the ARF-AID system at a safe harbor locus**. A) A step-wise strategy to integrate eGFP-ARF and TIR expressed from a CMV promoter. CRISPR is used to target the multi-cistronic construct into the AAVS1 locus. Primers F and R1 generate a PCR amplicon of 7256bp if the construct is inserted and 1892bp amplicon if the construct is not inserted. R2 primer recognizes a sequence internal to the TIR1 gene, so the primer combination F and R2 amplify only if the construct inserts into the AAVS1 locus. B) Clones show either heterozygous or homozygous integration of the insert. Compare the bands between the unintegrated (U) and integrated clones (HT -Heterozygous, HO -Homozygous). The clone in the fourth lane has one allele integrated and a deletion in the other allele. C) Western blotting confirms that the eGFP-ARF and TIR1 proteins are expressed.

## AID-ARF clamp system

All AID systems necessitate the expression of TIR1, so there are many progenitor cell lines and organisms that express TIR1 (Li et al., 2019; Natsume et al., 2016; Nishimura et al., 2009; Zhang et al., 2015; Holland et al., 2012). In an effort to repurpose these cell lines and organisms, but alleviate chronic degradation, we fused the AID-tag with ARF using a flexible linker to create the AID-ARF clamp (Figure 1) (Addgene #138174). We tagged ZNF143 with the AID-ARF clamp using a canonical TIR1 expressing progenitor cell line. Similar to the ARF-AID system, the AID-ARF clamp preserves near-endogenous protein expression (Figure 4A). Moreover, the AID-ARF clamp-tagged ZNF143 protein degraded rapidly upon auxin treatment (Figure 4B&C). Both ZNF143-AID-ARF clones tested showed an average half-life of 10 min (Figure 4C), which is similar to the reported ZNF143-AID degradation half-life of 11 min in the eGFP-ARF rescued system (Sathyan et al., 2019). In the future, we look forward to testing whether AID-ARF clamp-tagged proteins consistently degrade more rapidly as compared to the canonical AID and multi-cistronic ARF-AID systems.

Here we describe methods to implement the ARF-AID system in human cells, which is easily adaptable to other types of cells and for the AtAFB2 system. We will outline a protocol for 1) generation of ARF-TIR1 progenitor cells, 2) tagging the protein of interest with AID, and 3) testing degradation in the tagged cell lines.

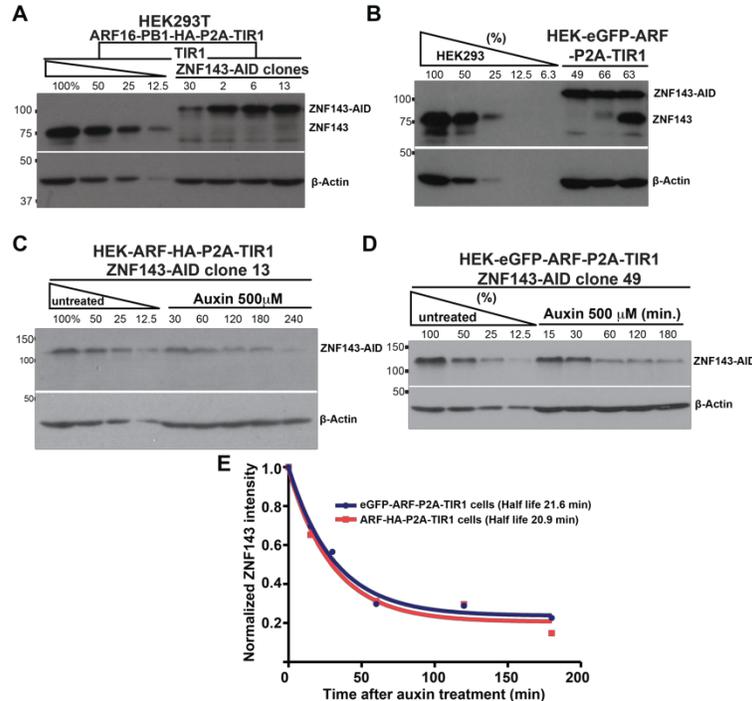

Figure 3. **The ARF-AID system preserves the endogenous expression levels of AID-tagged proteins and facilitates auxin-inducible degradation**. Tagging of ZNF143 with AID in both an ARF-HA-P2A-TIR1 (A) and eGFP-ARF-P2A-TIR1 (B) progenitor line preserved comparable expression levels compared to the progenitor lines. C&D) Both ZNF143-AID cell lines facilitate auxin-inducible ZNF143-AID degradation with (E) protein half lives of approximately 20 minutes.

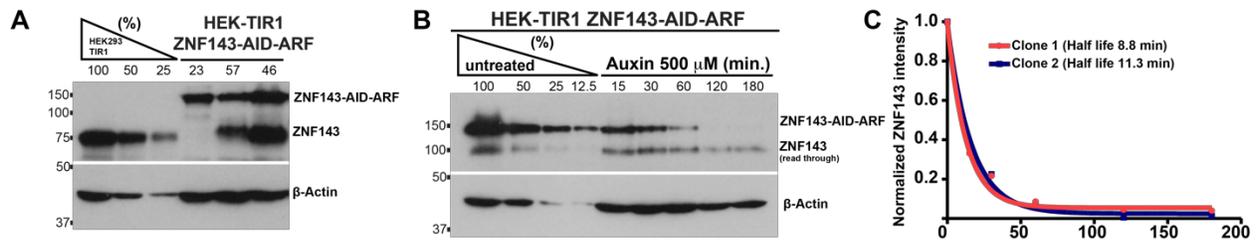

Figure 4. **The AID-ARF clamp system facilitates rapid auxin-inducible AID-ARF tagged proteins**. A) ZNF143-AID-ARF is expressed at comparable levels to the parental TIR1 expressing cells. B) ZNF143-AID-ARF is rapidly degraded upon addition of auxin with (C) protein half lives of approximately 10 minutes as assessed by two independent clones.

## STRATEGIC PLANNING

Establishing the ARF-AID system can be time consuming if all steps are performed serially. To reduce time, we recommend developing the ARF-TIR1 progenitor cells in parallel with testing gene-specific sgRNAs and PCR-amplifying the homology-directed repair (HDR) constructs (Figure 5).

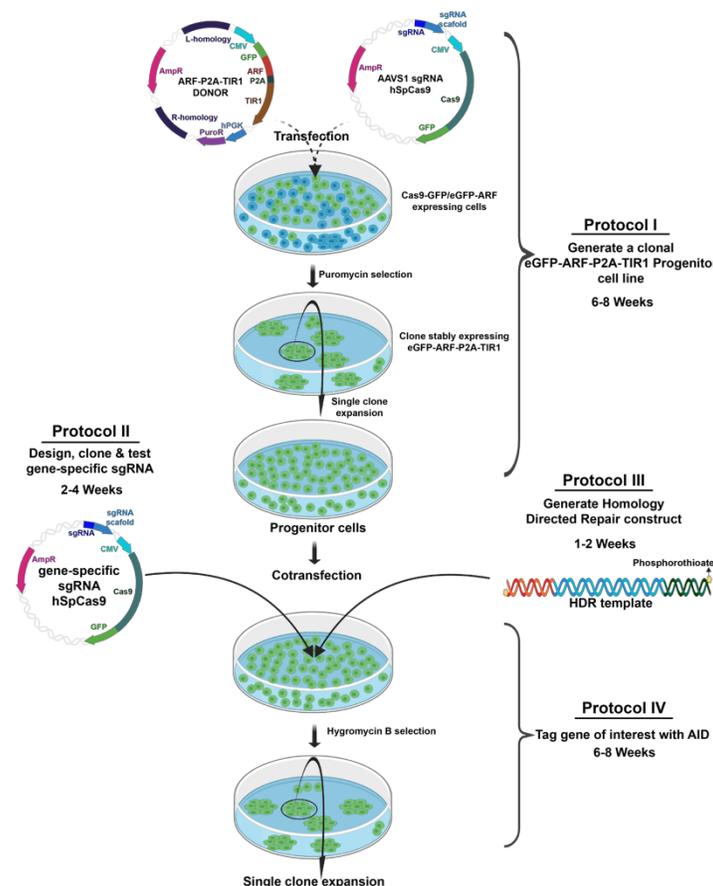

Figure 5. **The ARF-AID system has four distinct protocols**. Protocol I is used to generate a progenitor cell line that expresses all necessary components of the plant AID system. Protocols I, II, and III can be performed simultaneously. Protocol II involves testing sgRNA cleavage at the gene of interest and Protocol III outlines the design of the homology repair construct. Protocol IV incorporates all these components to tag the protein of interest with AID in the progenitor cell line.

*BASIC PROTOCOL 1*

**GENERATION OF eGFP-ARF-P2A-TIR1 OR ARF-HA-P2A-TIR1 PROGENITOR CELLS**

The first procedure for implementing the ARF-AID system is to establish ARF-TIR1 progenitor cells, as shown in Figure 2A. All the plasmids for integrating ARF-TIR1 into the AAVS1 locus in human cells are available from Addgene. The choice of transfection method varies depending on the cell type; lipofectamine 3000 works efficiently for HEK293T cells.

*Materials:*

HEK293T cells (ATCC)
pMGS46 (ARF16-PB1-HA-P2A-OsTIR1) Addgene #126580
pMGS56 (GFP-ARF16-PB1-P2A-OsTIR1) Addgene #129668
pMGS7 (AAVS1sgRNA) Addgene #126582
Opti-MEM reduced serum Medium (Gibco 31985-070)
Lipofectamine 3000 (Invitrogen L3000-015)
DNeasy blood and tissue genomic DNA isolation kit (Qiagen 69504)
Platinum Taq DNA Polymerase High Fidelity (Invitrogen 11304011)
10 mM dNTP mix (Invitrogen 18427013)
SYBR Safe DNA Gel Stain (Invitrogen S33102)
Puromycin (Gibco A11138-03)
Cloning cylinder (Bel-Art SP Scienceware 378470100)
Nonfat dry milk (BioRad 1706404)
Anti-OsTIR1 pAb (MBL PD048)
Anti-GFP mAb (Sigma 11814460001)
Anti-HA11 mAb (1:1000; BioLegend 901513)
Anti-β-Actin mAb (1:5000; Sigma, A1978)
Trypsin 0.05% (Gibco 25300054)
PBS (Gibco 10010023)

*Reagents and Solutions*

**HEK293T growth media (565 ml)**

DMEM (Gibco 11965-092, 500 ml)
Sodium pyruvate (100mM, Gibco 11360070) 5 ml
L-Glutamine 100x (Gibco 25030081) 5 ml
Fetal bovine serum (VWR 89510186) 50 ml
Pen/Strep (Gibco 15140122) 5 ml

**5x SDS buffer** (Laemmli buffer) (stored at room temperature)
312.5mM Tris HCl pH 6.8
10% SDS

50% (w/v) glycerol
0.05% bromophenol blue

Dilute 5x SDS buffer with water to 2x and add 50 µl 2-mercaptoethanol per ml of 2x SDS sample buffer prior to use. The 2x SDS sample buffer with 2-mercaptoethanol can be stored at -20°C.

**50x Tris Acetate EDTA (TAE) buffer, pH 7.2**

2M Tris base
1M Sodium Acetate
50 mM EDTA
pH to 7.2 with acetic acid and store at room temperature. Make 1x working solution by diluting in double distilled water.

*Cotransfection of eGFP-ARF-P2A-TIR1 or ARF-HA-P2A-TIR1 and AAVS1 sgRNA:*

1-1   Remove media from the HEK293T cells and wash with PBS. Add 0.5 ml 0.05% trypsin to the plate. Incubate the cells for 2-3 minutes, rinse and collect cells with 10 ml of media.

1-2   Remove trypsin by centrifugation at 500xg for 5 minutes using a swinging bucket centrifuge, remove supernatant, and resuspend the cells in 10 ml fresh media.

1-3   Seed 2.0-3 x$10^5$ cells per well of a six-well plate to get 30 to 40% confluent cells the next day.

> *Cotransfect 1 µg each of AAVS1 sgRNA (Addgene # 126582) and eGFP-ARF-P2A-TIR1 (Addgene # 129668) or ARF-HA-P2A-TIR1 (Addgene # 126582) plasmids with lipofectamine 3000 reagents. Transfection reagents or methods of introducing foreign DNA into cells will vary by cell line. For HEK-293T cells we recommend a ratio of 2.5 µl of lipofectamine for every µg of DNA. As a negative control, transfect the parental vector pX458 with the repair plasmid and include one well as non-transfected control.*

1-5   Add 125 µl Opti-MEM I reduced serum media in two 1.5 ml tubes for each transfection; one tube is labeled with the description of the DNA sample and the other is labeled as Lipofectamine. Add 5 µl Lipofectamine 3000 reagent to the tube labeled Lipofectamine. Add 5 µl p3000 reagent to the tube labeled DNA sample.

1-6   Add 1 µg each of AAVS1 sgRNA (Addgene # *126582*) and eGFP-ARF-P2A-TIR1 (Addgene # *129668*) or ARF-HA-P2A-TIR1 (Addgene # *126580*) plasmids into the p3000 reagent mixture and mix by pipetting up and down three times. Transfect *parental vector pX458 with* eGFP-ARF-P2A-TIR1 or ARF-HA-P2A-TIR1 as negative control.

> *If the transfection efficiency is low, increase the amount of Lipofectamine 3000 and p3000 reagents to 7.5 µl.*

1-7   Transfer the lipofectamine mixture (step 1-5) to the DNA/p3000 mixture (step 1-6) dropwise slowly. This tube contains ~260 µl of transfection reagents, DNA and Opti-MEM. Mix by pipetting up and down three times with the same pipette tip.

1-8   Incubate at room temperature for 15 minutes and transfer the DNA complex into the cells dropwise in a serpentine pattern to uniformly distribute the reagent across the plate.

1-9   Rock the plate sideways four times and front to back four times to distribute the reagent even over the cells--do not swirl the plate. Place cells in the incubator.

1-10   Twenty-four hours after transfection, replace the media with fresh media.

1-11   Forty-eight hours after transfection expand each well of the six-well plate into a 10 cm plate with 10 ml media as described in steps 1-1 and 1-2 .

> *Expanding cells into 10 cm plates spreads colonies ensures that the cells are sufficiently sparse to allow clonal colonies to form in isolation. Transfect multiple wells to ensure that a sufficient number of colonies survive.*

1-12   Seventy-two hours after transfection, begin selection by adding 1 µg/ml puromycin. Antibiotic selection concentration varies between cell types. We recommend plotting a titration vs. cell viability curve to determine the lowest concentration at which nearly all cells die within 5 days.

> *If the number of colonies are low, begin selection after 120 hours, as opposed to 72. The addition of conditioned media (8:2 ratio of fresh media to filtered cultured media from the same cell type) increases the survival of the clones.*

1-13a   After three days of selection replace with fresh media containing puromycin (1µg/ml) and continue selection for two days.

1-13b   (optional) After five total days if there are many remaining cells in the negative control plate, then replace with fresh media containing puromycin and continue selection for two days for a total of seven days under selection.

1-14   After five to seven days under selection all the cells in the untransfected control well should be dead. Remove selection media and expand cells without selection.

> Allow colonies to grow and expand. Usually, it takes 2 to 3 weeks for colonies to appear after beginning the puromycin selection (Figure 2A&5), but the time frame is variable and dependent upon the cell line. Check the plate under a microscope and scan for colonies that appear after two weeks. The presence of a substantially higher number of colonies in the transfected plates compared to the control plasmid-transfected or untransfected plates suggests successful integration of the plasmid. The control sgRNA plasmid transfected plates often contain some colonies, but many fewer.

### *Picking colonies*

1-15   Mark individual colonies on the bottom of the plate using a marker. Check eGFP-ARF-P2A-TIR1 cells using a fluorescent microscope to identify GFP-positive colonies. Select colonies in which all the cells in the colony exhibit uniform nuclear expression of GFP. If the plasmid is modified to localize ARF into cytoplasm or uniformly throughout the cell, then mark those with appropriate GFP distribution. In the case of the ARF-HA-P2A-TIR1 transfected cells select all colonies.

1-16   Carefully pick individual colonies by taking 10 µl 0.05% trypsin solution using a P20 pipette set to 20 µl. Alternatively, use cloning cylinders to pick individual colonies.

> For hand picking colonies without the use of cloning cylinders: 1) hold the plate at a 45-degree angle to pool the media away from the colonies; 2) use a P20 pipette set to 20 µl and aspirate ~10 µl of trypsin solution into the pipette tip; 3) dispense a few microliters of the trypsin solution onto the colony such that a small droplet is formed between the plate and the pipette tip, taking care not to dispense so much that the droplet rolls down the plate; 4) scrape the colony with the pipette tip while the trypsin media remains as a bridge between the plate and the tip; and 5) once the colony is dislodged, aspirate the cells into the tip and transfer of the entire colony to the 96 well plate. Resuspend the colonies in 200 µl media and transfer into 24 well plates containing 1 ml media per well.

> *Using cloning cylinders: 1) dispense silicone grease onto a glass petri dish and autoclave it along with a pair of forceps; 2) aspirate media from the plate and wash with PBS; 3) hold the plate at a 45-degree angle to pool the PBS away from the colonies; 4) use forceps to pick up a cloning cylinder, dip the thicker edge into silicone grease, and place it over the colony; 5) add 20 µl of trypsin solution and incubate at 37° until cells begin to detach; 6) resuspend the colony in 100 µl media and transfer into a 24 well plate containing 1 ml media per well.*

1-17   When the cells reach confluency continue to the next step.

### *Expanding cells, screening for genomic integration of the ARF-TIR1, and Western blotting*

1-18   Collect cells by pipetting up and down and transfer 100 µl into a new plate containing media to continue passaging the cells. Transfer the remaining ~0.9 mls into a 1.5 ml tube.

> *The plate allows the expansion of the positive colonies. HEK293T cells attach loosely to the plate, so pipetting is enough to dislodge the cells. For other cell lines, collect cells by trypsinization.*

1-19   Aliquot 100 µl of the 0.9 mls within the eppendorf tube into a fresh tube for genomic DNA isolation and PCR. Place cells immediately on ice.

1-20   Centrifuge the remaining 0.8 mls of cells at 6000 x g for 2 minutes using a fixed angle rotor table top centrifuge and remove media.

1-21   Add 100 µl 2x SDS sample buffer (Laemmli buffer) and mix thoroughly by pipetting up and down to generate cell lysate for Western blotting.

1-22   Heat the lysate at 95°C for 5 minutes on a heating block and vortex for 10 seconds. Place back on the heating block for another 5 minutes. Briefly centrifuge the samples at 5000xg and store at -20°C.

### *Screening genomic integration at the AAVS1 locus*

To screen the integration of ARF-TIR1 at the AAVS1 locus, we used genomic PCR using primers that amplify the integrated plasmid DNA.

***Genomic DNA isolation***

1-23   Centrifuge cells from step 1-21 at 6000 x g for 2 minutes using a fixed angle rotor table top centrifuge, remove media, and either flash freeze or proceed immediately with gDNA isolation.

> *Use any genomic DNA isolation kit or conventional phenol-chloroform isolation of DNA to isolate DNA from the cells. Here we used the Qiagen genomic DNA isolation kit. The method described here is adapted from the kit manual with minor changes (DNeasy Blood and Tissue Kit, 69504).*

1-24   Resuspend the cells in 200 µl PBS and add 20 µl proteinase K.

1-25   Lyse cells by adding 200 µl Buffer AL and mix thoroughly by vortexing to form a homogenous lysate.

1-26   Incubate samples at 56°C for 10 minutes.

1-27   Add 200 µl 100% ethanol and vortex.

1-28   Transfer the lysate into a DNeasy Mini spin column placed in a 2 ml collection tube.

1-29   Spin at max speed in a tabletop centrifuge (10,000-17,000 x g) for 1 min and discard the flow-through.

1-30   Place the spin column back into the collection tube and add 500 µl Buffer AW1. Spin at 10,000-17,000 x g and discard the flow-through.

1-31   Place the spin column back into the collection tube and wash by adding 500 µl Buffer AW2 and centrifuging at 10,000-17000 x g for 1 min.

1-32   Discard the flow-through and centrifuge again at maximum speed for 2 min.

1-33   Place the spin column into a 1.5 ml tube and add 100 µl nuclease-free water to the center of the column. Spin at 9000 x g for 2 min.

> *Centrifuging at 9000 x g reduces the chance of breaking off the Eppendorf tube's lid.*

1-34   Quantify DNA using NanoDrop and store at -20°C.

***Genomic DNA PCR***

A wide variety of Taq DNA polymerase products are available for PCR. Each may need a slightly different PCR condition. We recommend Platinum Taq DNA Polymerase High Fidelity for validating ARF-TIR1 genomic DNA insertion using PCR. The forward and reverse primers flank the genomic DNA integration site of the ARF-TIR1 construct (Figure 2A). If there is no insert, the PCR produces a smaller product with the flanking primer, but yields a larger product if ARF-TIR1 is properly integrated (details below). Heterozygous integration of the construct results in the two bands. Independently, we also perform a PCR using the flanking forward primer and a primer that is internal to the insert to confirm the integration of the construct at the locus (Figure 2B).

1-35 Make a PCR master mix by adding all the components except genomic DNA for the required number of reactions:

PCR reaction mix for one sample:

| | |
|---|---|
| Genomic DNA (10ng/µl) | 5.0 µl |
| 10 µM Primer F | 1.0 µl |
| 10 µM Primer R or R 2 | 1.0 µl |
| 10X High Fidelity buffer | 2.5 µl |
| DMSO | 0.5 µl |
| 10 mM DNTP | 1.0 µl |
| 50 mM MgSO$_4$ | 1.0 µl |
| Platinum Taq DNA Polymerase High Fidelity | 0.5 µl |
| Water | 12.5 µl |

1-36 Aliquot 20 µl of the master mix into 0.2 ml PCR tubes and add 5 µl of 10ng/µl genomic DNA into each reaction mix. Use the parental HEK293T DNA as a negative control.

**PCR condition as follows:**

Initial denaturation
95°C 5 min

30 cycles:
95°C 30sec
59°C 30 sec
68°C 7 min

Final extension
68°C 10 min
4°C hold ∞

**Primers used**

AAVS1GenomicF (F)
5'-CTGCCGTCTCTCTCCTGAGT-3'
AAVS1GenomicR (R)
5'-ACAGTTGGAGGAGAATCCACC-3'
Internal Primer (R2)
5'-ATTATGATCTAGAGTCGCGGC-3'

**Electrophoresis**

1-37   Make 1% agarose gel with 1XTAE buffer.

1-38   Add 5 µl 6x DNA sample buffer to the PCR tubes and load onto the 1% agarose gel.

1-39   Run the samples at constant 90V for 60 minutes.

1-40   Stain the gel with SYBR Safe DNA gel stain diluted at 1:10000 with 1xTAE buffer for 10 minutes and wash twice with 1xTAE for 10 minutes each.

1-41   Visualize the bands using a UV transilluminator (Figure 2A&B).

> *The negative control produces an amplicon of 1892bp with genomic F and R primers. Successful integration of the construct homozygously produces an amplicon of 7256bp, and heterozygous integration shows both amplicons. The PCR using the AAVS1GenomicF (F) and Internal Primer (R2) primers produce a 4677 bp amplicon only in the integrated cells, not in the negative and control cells* (Figure 2A&B).

**Confirmation of the clones by Western blotting**

*Integration of the ARF and TIR1 genes at the AAVS1 locus does not necessarily mean the genes are expressed. Use Western blotting to test whether ARF and TIR1 proteins are present in the cell.*

1-42   Select homozygously integrated ARF/TIR1 clones from the PCR screen.

1-43   Thaw the frozen protein lysate (Step 1-24) and heat again at 95°C for 3 to 5 minutes.

1-44   Separate proteins by loading 10 µl of the samples on a 10% acrylamide gel, transfer the proteins onto nitrocellulose or PVDF membrane, block membrane with 7.5% nonfat dry milk, and probe with anti-GFP or anti-HA and anti-TIR1 antibodies. β-Actin (1:5000; Sigma, A1978) can be used as a loading control. Include parental cell lysate as a negative control (Figure 2C).

After insertion and expression are confirmed by genomic PCR and Western blot, the construct should be sequenced to confirm that mutations were not incorporated during the process. This cell line will serve as the progenitor cell line for tagging genes of interest with full-length AID.

## BASIC PROTOCOL 2

### TAGGING A GENE OF INTEREST WITH AID

The next step in developing the ARF-AID system is to tag the gene of interest with AID. The ARF-AID system requires full-length AID (Figure 1) because the characterized interaction domains of AID with ARF are domains III and IV. Domains I and II are involved in the interaction with TIR1. The mini-AID lacks domains III and IV and will not interact with ARF to stabilize the protein in the absence of auxin (Sathyan et al., 2019). Note that the antibiotic selection marker (HygroR) is co-transcribed with AID and the protein products are separated during translation. Therefore, the resistance marker will be expressed at levels comparable to the target protein.

### *Materials:*

HEK293T ARF-P2A-TIR1 progenitor cells
pMGS54 (AID-P2A-Hygromycin) Addgene #126583
pMGS58 (Hygromycin-P2A-AID) Addgene #135311
Platinum Taq DNA Polymerase (Invitrogen 10966-034)
QIAprep Spin miniprep kit (Qiagen 27104)
QIAquick gel extraction kit (Qiagen 28704)
BbsI (New England Biolabs R0539S)
T4 DNA ligase (New England Biolabs)
T4 Polynucleotide kinase (New England Biolabs M6201S)

Max efficiency DH5a (Invitrogen 18258-012)
Surveyor kit (IDT 706025)
sgRNA sequencing primers
    LKO-1 5' primer 5'-GACTATCATATGCTTACCGT-3'
    U6 promoter primer 5'-CACAAAGATATTAGTACAAAATACG-3'
Carbenicillin (Disodium) (Goldbio C-103-25)
LB plate
100X BSA (New England Biolabs B9001S)
Hygromycin B (Invitrogen 10687010)
LB broth Media (Fisher, BP1426)
LB agar (Fisher, BP1425)
S.O.C medium (Invitrogen 15544034)
Betaine solution (Sigma B0300)

| | |
|---|---|
| FOXM1 sgRNA1 | 5'-GCAGGGCTCTACTGTAGCTC-3' |
| FOXM1 sgRNA2 | 5'-GGGACCAGTTGATGTTGTCA-3' |
| FOXM1 forward primer | 5'-TCTGGCAGTCTCTGGATAATGAT-3' |
| FOXM1 reverse primer | 5'-GCTGATGGATCTCAGCACCACTC-3' |

*Design gene-specific sgRNA and cloning*

Adding any tag to the N-terminal or C-terminal of the protein could disrupt the function of a gene. Therefore, the functionality of the AID-tagged proteins should be empirically determined. We recommend designing sgRNAs to both ends and then testing the functionality of both the N and C-terminally tagged proteins. There are several free tools available to design sgRNA, such as from Benchling, CHOPCHOP, E-CRISP, and CRISPOR (Labun et al., 2019; Heigwer et al., 2014; Benchling, 2019; Concordet and Haeussler, 2018). We outline the process of using Benchling or manually choosing sgRNAs based on the presence of a PAM sequence within 25 bases of the start or stop codon.

    Consider two parameters while choosing an sgRNA: the distance from the desired homologous repair site and the specificity and off-target effects of the sgRNA. The efficiency of the homologous repair of the AID-tag at the cut site increases if the required homologous repair site is near the sgRNA cut site (O'Brien et al., 2019; Inui et al., 2014). Increased distance between the cut site and the start codon or the stop codon can create challenges when designing homology arms. For example, when inserting an N-terminal tag and using a guide with a cut site upstream of the start codon, the downstream homology arm can only extend to the start codon. The upstream homology arm can either end at the cut site, removing the 5' UTR from the resulting

product, or continue to the start codon. If this homology arm has extensive homology with both sides of the cut site, then the cut may be repaired without proper insertion of the template. The same types of challenges occur with cut sites internal to the protein coding region, as well as with C-terminal tagging. However, silent mutations can be introduced into the homology arms to decrease sequence homology within the protein coding region without reestablishing a sgRNA recognition site.

The sgRNA's proximity to the homologous recombination site takes priority over the specificity scores. Design at least three sgRNA to each terminus and clone into pX458 (Ran et al., 2013) (or an appropriate vector), which harbors GFP that can be used to quantify the efficiency of transfection. If an ideal sgRNA is not found, check the possibility of using other Cas9 enzymes with different PAM sequence requirements (Kleinstiver et al., 2015). Guide RNAs must be designed with the 3' PAM sequence but do not include the PAM sequence in the cloned sgRNA construct.

The U6 promoter in the pSpCas9(BB)-2A-GFP (pX458) plasmid requires a 'G' at the beginning of the guide to efficiently transcribe the sgRNA. If the sgRNA designed does not have a 'G' at the 5'-end, add one 'G' at the 5' of the forward sequence of the guide RNA and reverse complement of the 'G' in the reverse sequence. A general strategy to clone and test the efficiency of gene-specific sgRNA is shown in Figure 6.

### *Designing sgRNA using Benchling*

#### **Importing target gene sequence**

1. From the left navigation bar in Benchling, click Create > CRISPR > CRISPR Guides.

2. Search for the target gene by gene ID or name, and choose the appropriate genome assembly and transcript.

*Optional: if there are too few nucleotides imported upstream of the start codon or downstream of the stop codon for the desired homology arm length, choose "Show Advanced Options" to import additional nucleotides.*

3. Guide parameters can be left at the default settings or adjusted as needed.

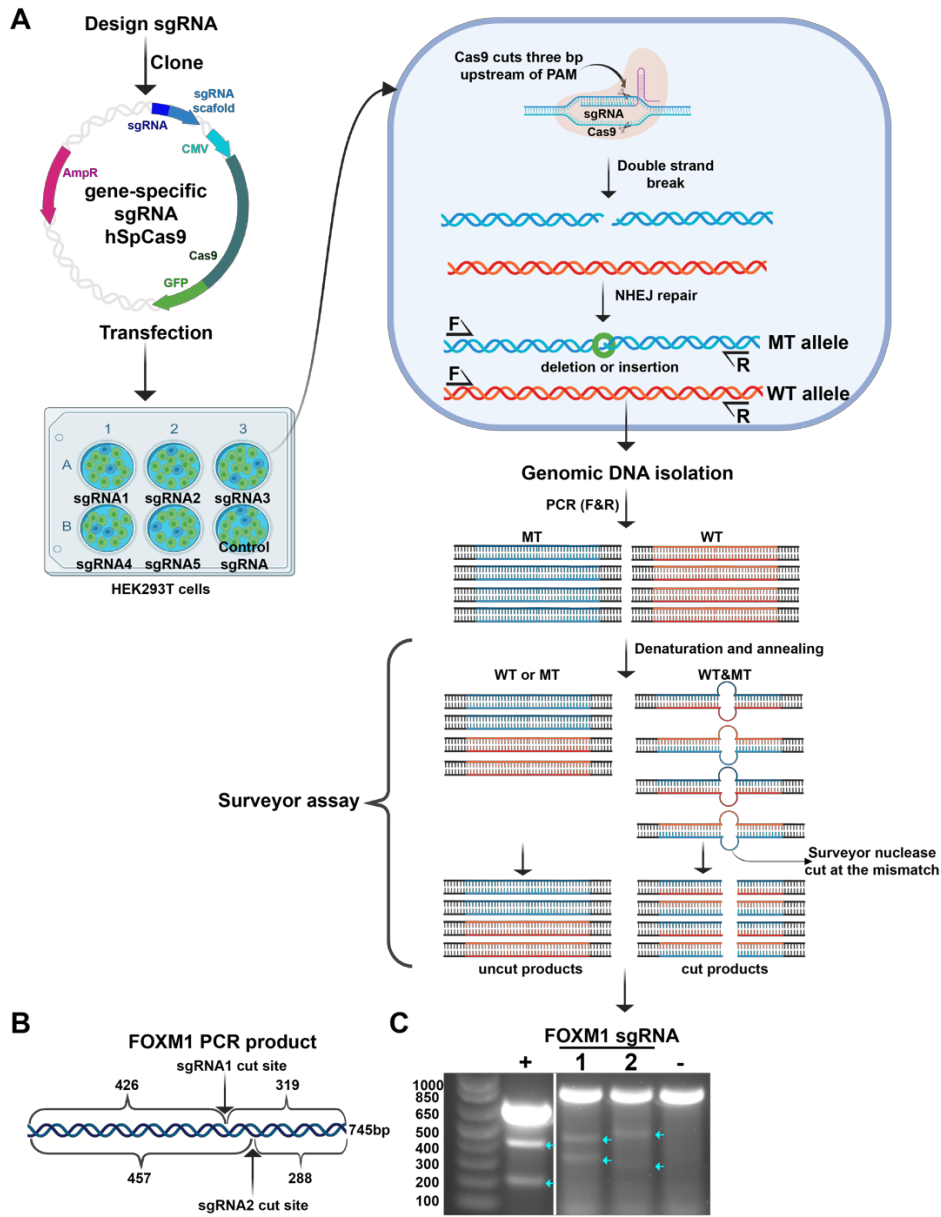

Figure 6. **Design and test gene-specific sgRNA**. A) A Surveyor assay is used to detect the mutation generated by non-homologous end joining (green circle) at the sgRNA targeted sites. B) Two sgRNAs were designed proximal to the 3' end of FOXM1 gene. PCR primers flank the sgRNA cutting sites and the sizes of Surveyor-cleaved products are indicated. C) The PCR product from two FOXM1 sgRNA transfected HEK293T cells were digested with surveyor and visualized. The positive control is provided in the surveyor assay kit, and the negative control is a PCR product from HEK293T cells without an sgRNA transfection. Cyan arrows indicate the bands that result from heteroduplex digestion.

### Choosing guides

1. Highlight the region 25 nucleotides upstream and 25 nucleotides downstream around the start codon (for N-terminal tagging) or the stop codon (for C-terminal tagging) and click "Create" to create a target sequence.

2. Benchling provides a list of guides targeting the region, along with their predicted On-Target Scores (Doench et al., 2016) and Off-Target Scores (Hsu et al., 2013). In addition to maximizing these scores, distance to the start or stop codon should be minimized (see above).

3. For guides chosen, click "Assemble" and choose the vector into which they will be cloned, e.g. pX458.

4. For each guide, the resulting Assembly shows the predicted plasmid after cloning.

5. Return to the tab with the imported sequence, click "copy the primer list", and paste the sequences for all generated oligonucleotides into a spreadsheet.

*The two oligonucleotides (FWD and REV) can be synthesized for downstream cloning.*

Alternatively, choose the 20 nucleotide target sequence without the PAM and add overhangs to clone into pX458. If the sequence does not start with a G, add a G to the 5'-end. Append 5'-CACC-3' to the 5'-end of the forward sequence and 5'-AAAC-3' to the 5'-end of the reverse complement of the target sequence, ensuring that the 3' base is the complementing C nucleotide (Cong et al., 2013).

*Cloning the guide RNA into pSpCas9(BB)-2A-GFP (pX458)*

**Digestion of the vector with BbsI**

2-1   Digest 3 µg of pX458 with BbsI overnight at 37°C.

*The overnight digestion with BbsI significantly reduces negative colonies.*

**BbsI reaction mix:**

| | |
|---|---|
| NEB buffer 2.1 | 5.0 µl |
| BbsI enzyme | 1.0 µl |
| 100X BSA | 0.5 µl |
| pX458 plasmid | 3.0 µg |
| Water | to 50 µl |

### *Gel purification of the digested plasmid*

2-2  Add 8 µl loading dye to the mixture and load onto 1% agarose gel. Use the same amount of the undigested pX458 vector as a negative control.

2-3  Use a transilluminator to excise the digested band using a clean scalpel and use the QIAquick gel extraction kit (Qiagen, 28704) to isolate the digested plasmid.

> *The use of the blue light dark reader transilluminator (Clare Chemical Research, USA) may reduce DNA damage during gel excision. The digested plasmid runs between supercoiled and relaxed DNA in the undigested control sample.*

### *DNA purification*

*The method described here is adapted from the kit manual with minor changes (QIAquick gel extraction kit, Qiagen, 28704).*

2-4  Add three volumes of buffer QG to one volume of gel; consider 1 mg of gel equivalent to 1 µl. For example, add 300 µl of QG buffer for 100 mg of gel.

2-5  Incubate at 50°C for 10 minutes. Mix by vortexing every three minutes to dissolve the gel. Once the gel has completely dissolved, add 10 µl of 3M sodium acetate, pH 5.0, irrespective of the color of the mixture, and briefly mix by vortexing.

> *The addition of sodium acetate significantly increases the final yield of the DNA.*

2-6  Add one gel-volume of isopropanol and vortex briefly.

2-7    Transfer the mixture into the QIAquick column and centrifuge at 17000 x g for one minute.

> *If there is more dissolved gel mixture left, add to the same column and repeat step 2-7.*

2-8    Remove the flow-through and place the column back to the collection tube. Add 500 µl QG buffer to the column and centrifuge at 17000 x g for one minute.

2-9    Remove the flow-through and place the column back to the collection tube. To wash the column, add 750 µl of PE buffer and centrifuge at 17000 x g for one minute.

2-10   Discard flow-through and replace the column into the same tube and centrifuge again at 17000 x g for 2 minutes.

2-11   Place the column into a new clean 1.5 ml Eppendorf tube and add 50 µl DNase and RNase free water. Incubate for two minutes and centrifuge at 9000 x g for two minutes.

> *Centrifuging at 9000 x g reduces the chance of breaking off the Eppendorf tube's lid.*

2-12   Quantify DNA using a NanoDrop and store at -20°C.

***Ligation of guide sequence into pX458 plasmid***

2-13   Make 100 µM solution of the forward and reverse strands of the sgRNA guide sequence in nuclease free water.

2-14   Phosphorylate and anneal the forward and reverse oligonucleotides of the sgRNA guide sequence in one reaction.

**Phosphorylation and annealing reaction mix:**

| | |
|---|---|
| Forward (100 µM) | 1 µl |
| Reverse (100 µM) | 1 µl |
| T4 DNA Ligase buffer | 1 µl |
| T4 PNK | 1 µl |
| Water | 6 µl |

Incubate in a thermocycler
37°C 30 min
95°C 5 min
Ramp down to 4°C at 4°C/min rate.

2-15  Dilute the annealed sgRNA guide sequence pair 20x by adding 190 µl of nuclease-free water. We use one microliter of this 500nM dsDNA product for ligation.

**DNA ligation mix:**

| | |
|---|---|
| BbsI digested pX458 | 50 ng |
| annealed sgRNA | 1 µl |
| T4 ligase Buffer | 1 µl |
| T4 DNA Ligase | 1 µl |
| Water | to 10 µl |

2-16  Incubate ligation mix for 2 hours at room temperature or 16°C overnight. Include no insert control to see the rate of negative colonies.

> *Typically, we incubate 2 hours at room temperature and start the transformation. Keep the rest of the ligation mix at 16°C overnight. If the first transformation does not result in colonies, then retransform the ligated product the next day.*

2-17  Add 2 µl of the ligated product to 20 µl chemically competent E. coli. We used Max efficiency DH5a competent cells from Invitrogen for transformation.

2-18  Incubate on ice for 30 min and then heat shock for 30 seconds at 42°C.

2-19  Put back on ice for 2 minutes.

2-20  Add 250 µl LB or SOC media and incubate at 37°C for one hour.

2-21  Plate the cells (272 µl) on a carbenicillin plate (stable version of ampicillin) and incubate at 37°C overnight.

2-22  Pick three colonies from each plate and inoculate in LB containing carbenicillin and incubate overnight at 37°C.

2-23   Pellet the bacterial culture by centrifugation at 17000 x g for one minute and follow the steps in the QIAprep spin miniprep kit. Elute DNA in 50 µl nuclease free water.

2-24   Confirm the sgRNA insertion by sequencing using the LKO.1 5' primer or the U6 promoter primer.

*Testing sgRNA using surveyor assay*

**Transfection of the sgRNA construct**

2-25   Plate 30% confluent HEK293T cells in 6 well plate (start with $2 \times 10^5$ cells).

2-26   Transfect 1 µg of sgRNA plasmid with lipofectamine 3000 reagents as described in section 1-5 - 1-9.

2-27   Twenty-four hours after transfection, replace media with 2 ml new media.

> *GFP in the pX458 plasmid allows easy assessment of the transfection efficiency. Observe the transfected cells under a fluorescent microscope using a green filter. Higher transfection rates make it easier to assess the efficacy of the sgRNA using surveyor assay*

2-28   Seventy-two hours after transfection, remove 1 ml of media and collect the cells in the remaining 1 ml of media by pipetting up and down and transfer into a 1.5 ml tube.

2-29   Centrifuge cells at 6000 x g for 2 minutes using a fixed angle rotor table top centrifuge and remove media. Cells can be stored at -20°C or directly begin genomic DNA isolation.

2-30   Proceed with genomic DNA isolation, as described in sections 1-24 – 1-34.

**Genomic DNA PCR**

**Primer design**

To amplify the region targeted by the sgRNA, design a set of forward and reverse primers. The forward primer should be ~250bp upstream from the sgRNA cutting site and the reverse primer should be ~250bp downstream. To facilitate mutation detection, we recommend an amplicon size of 500bp to 1Kb with an sgRNA cut site close to the center of the amplicon. There are several tools to pick appropriate primer sets, such as from Primer3, IDT or Benchling (Untergasser et al., 2012; Benchling, 2019) https://www.idtdna.com/pages/tools/primerquest). The double stranded DNA breaks generated by CRISPR-Cas9 are typically repaired by error prone non-homologous repair, which results in several types of mutations. The mutated PCR products, when annealed with the wild type or other mutated PCR products, generate a mismatch proximal to the cut site. The surveyor enzymes recognize this mismatch and cleave the heteroduplex at the site of mismatch (Figure 6). The assay produces three fragments, one undigested and two digested fragments. Depending on the site of mismatch, the digested fragments run either as one or two bands (Figure 6B&C). If the sgRNA cutting site is at the center of the PCR product, it generates two unresolvable fragments, whereas unequal fragments run as distinct bands. There is no need to mix wild type PCR products with the PCR products of the sgRNA transfected cells for the assay because many different repair products will form and other sites in the population will remain unmodified. One set of primers is required for each terminus of the gene to test sgRNAs that target these sites.

We used Platinum Taq DNA polymerase or Platinum Taq DNA Polymerase High Fidelity for genomic PCR. For each primer set, it is necessary to determine the annealing temperature empirically. We find that an annealing temperature within 3 degrees of the lowest melting temperature (Tm) works well. Additionally, adding denaturants such as DMSO or Betaine may help to amplify GC rich genomic regions.

2-31   Make a PCR master mix by adding all the components except genomic DNA and gene specific primers adjusted for the total number of reactions.

2-32   Aliquot 18 µl of the master mix into 0.2 ml PCR tubes and add 5 µl of 10 ng/µl genomic DNA into each reaction mix. The parental HEK293T DNA serves as a negative control.

2-33   Add 1.0 µl each of forward and reverse primers of the corresponding gene targeted.

2-34   **PCR mix:**

| | |
|---|---|
| Genomic DNA | 5.0 µl |
| Gene Primer F | 1.0 µl |
| Gene Primer R | 1.0 µl |
| 10X HF buffer | 2.5 µl |
| DMSO (100%) | 0.5 µl |
| 10mM DNTP | 1.0 µl |
| MgSO$_4$ | 1.0 µl |
| Platinum Taq DNA Polymerase High Fidelity | 0.5 µl |
| Water | 12.5 µl |

*Annealing and extension of the PCR steps depends on the primers and Taq DNA polymerase used.*

**Run PCR as follows:**

Initial denaturation
95°C 5 min

30 cycles of
95°C 30sec
___°C. 30 sec (Annealing temperature changes based on the primers used)
68°C 1 min

Final extension
68°C 10 min
4°C hold ∞

2-35   Check PCR amplification by running 5 µl of the products on an agarose gel. Add 2 µl 6x DNA sample buffer to 5 µl of PCR product and load onto a 1% agarose gel.

*Keep the remainder of the PCR product for the surveyor assay. This step makes sure the PCR worked before starting the surveyor assay.*

2-36   Run the samples at constant 90V and visualize the bands using a UV transilluminator (follow steps 1-37 - 1-41). If there is only one bright PCR product, proceed with the surveyor assay.

**Surveyor assay**

This step involves denaturation followed by annealing of the PCR products to form heteroduplexes, followed by a surveyor nuclease reaction. The denaturation and renaturation step is essential because the final cycle of PCR generates homoduplexes that are not recognized by the surveyor nuclease.

2-37   Denature the PCR products from step 2-34 at 95°C for 10 minutes and then allow to renature stepwise using the following program:

    95°C    10 min
    85°C     1 min
    75°C     1 min
    65°C     1 min
    55°C     1 min
    45°C     1 min
    35°C     1 min
    25°C     1 min
    4°C hold ∞
    A ramp down rate of 0.3°C/sec is recommended.

2-38   **Surveyor nuclease reaction mix:**

    Reannealed PCR product        20.0 µl
    Surveyor Nuclease S            1.0 µl
    Surveyor Enhancer S            1.0 µl
    0.15 M MgCl2 Solution          2.0 µl

2-39   Mix by pipetting and incubate at 42°C for 60 min in a thermocycler.

2-40   Stop the reaction by adding 2.4 µl of Stop Solution and mix.

> *Either directly electrophoresis the sample on a 2% agarose gel after mixing with loading dye or store at -20°C for future use.*

2-41   Run the samples at constant 90V and visualize the bands using a UV transilluminator. Follow steps 1-37 to 1-41 for running and visualizing the samples. The ratio of the undigested band with that of the digested band gives an estimate of relative sgRNA efficiency.

*As an example, we designed two sgRNA that target the 3' end of FOXM1 coding regions (Figure 6B&C).*

*Homology directed repair construct design*

Homology-Directed Repair (HDR) is the mechanism by which AID is translationally fused to the N or C terminus of the target gene. CRISPR-Cas9 is directed to the region by sgRNA and this complex cleaves double-stranded DNA, which can be repaired by non-homologous end joining or homologous recombination/repair. The presence of a repair construct is necessary to increase the probability of homologous repair. We designed repair constructs specific for the N-terminal and C-terminal of the protein (Figure 7). For C-terminal fusion, the repair construct consists of AID separated by a porcine teschovirus-1 ribosomal skipping sequence (P2A) (Kim et al., 2011) from the hygromycin resistance gene. In the N-terminal repair construct, the order is reversed: HygR-P2A-AID. In both cases, the AID is separated from the protein of interest by adding a linker sequence of 6-9 amino acids (3x GGS).

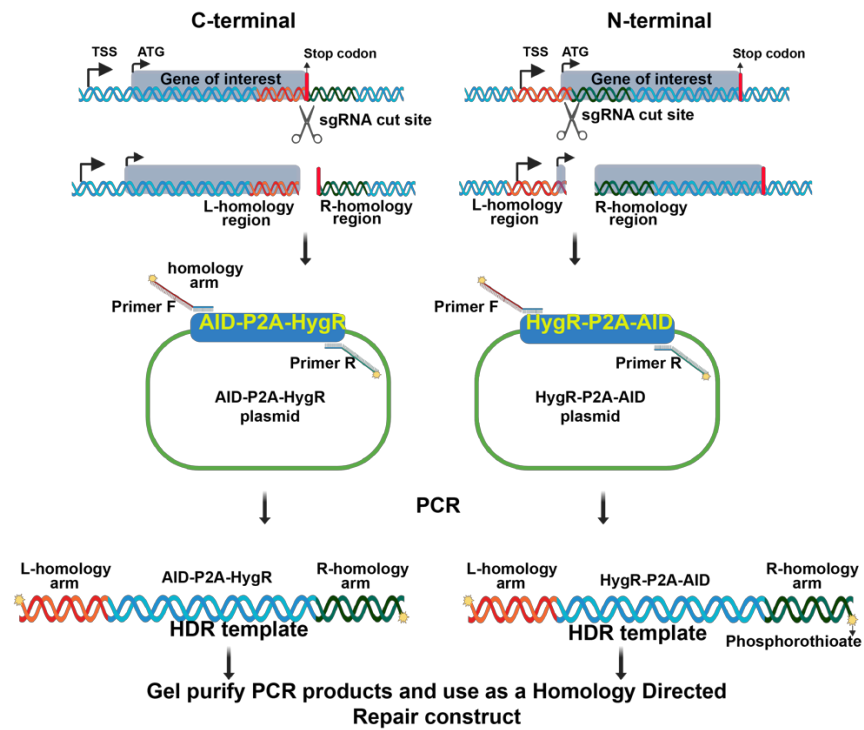

Figure 7. **The strategy for designing the Homology Directed Repair (HDR) construct for both the N and C termini**. A 50 nucleotide homology arm tail is added to the primers that amplify the HygR-P2A-AID (N) and AID-P2A-HygR (C) cassettes. PCR products are gel purified and used as an HDR template.

The HDR construct can be PCR products or cloned into a plasmid. If PCR products are used, two phosphorothioate moieties are added to the first two 5' nucleotides of each primer to increase PCR product stability in the cell (Zheng et al., 2014). Homology arms can vary in length from eight hundred nucleotides to less than 10 nucleotides (Sakuma et al., 2016; Lambrus et al., 2018; Paix et al., 2017). We recommend 50 nucleotide homology arms on both sides of the cut site for the AID integration (Sathyan et al., 2019).

In the C-terminal region, if the cut site is before the stop codon, we include wobble substitutions for C-terminal amino acids in the forward primer. These wobble substitutions have a two-fold purpose: 1) they ensure that the only homologous region in the donor is upstream of the cut site and 2) they prevent reconstitution of the sgRNA recognition site. Similarly, in the N-terminal region, if the cut site is after the start codon, we include wobble substitutions for N-terminal amino acids in the reverse primer. Simulate HDR reaction by *in silico* PCR using the SnapGene software or other software. Confirm that the sgRNA recognition sequence is not recreated after the HDR and only the desired homologous sequences are available for recombination.

**Designing Primers for the C-terminal tagging of the protein**

The 3' UTR can be critical in the regulation of the gene expression, and any changes in the sequence could modulate expression levels. If possible, select sgRNAs that cut inside the gene before the stop codon. If the sgRNA cuts after the stop codon, choose the closest to the stop codon. This reduces the challenges affecting the regulatory elements of the gene while designing the repair construct. Consider using another CRISPR enzyme with a different PAM recognition sequence if there is no optimal sgRNA sequence available for the CRISPR-Cas9 system (Kleinstiver et al., 2015).

**Designing homology arm primers**

In all cases in which the desired homologous recombination event recreates the original guide sequence with fewer than two mismatches and an intact PAM sequence, mutate the relevant homology arm to abrogate guide binding. Use a silent mutation to destroy the PAM sequence if possible, or use two silent mutations near the 3' end of the guide sequence (Cong et al., 2013). Check the evolutionary conservation of the wobble nucleotides (Ramani et al., 2019) to prioritize less conserved nucleotides. Check a codon usage chart to prioritize codons used at a similar frequency to the replaced codon (Athey et al., 2017).

**Upstream homology arm (coding strand primer design)**

### *Cut site upstream of the stop codon*

The upstream homology arm begins 50 bases upstream of the cut site and the last base is the nucleotide immediately upstream of the first stop codon base. The coding nucleotides downstream of the cut site will need to be modified at their wobble bases.

Append the sequence given below so that the primer anneals to the AID-P2A-HygR cassette in pMGS54.

*Critical: confirm that the AID-P2A-HygR cassette is in frame with the protein after repair by in silico PCR using SnapGene or any other program.*

### *Cut site downstream of the stop codon*

The upstream homology arm begins 50 bases upstream of the cut site and the last base is the nucleotide immediately upstream of the first stop codon base.

Append 5'-GGTGGATCTGGAGGTTCAGGTGGCAGTGTCGAGCTGAATCT-3' to the 3'-end of the upstream homology arm for C-terminal tagging using the insert from pMGS54. This sequence contains a flexible linker region prior to the AID coding sequence.

*Critical: confirm that the AID-P2A-HygR cassette is in frame with the protein after repair by in silico PCR using the SnapGene or any other program.*

### **Downstream homology arm (template strand primer design)**

The downstream homology arm begins at the cut site and extends 50 bases downstream of the stop codon.

In the case of sgRNA cut site downstream of stop codon, the homology arm can extend to immediately upstream of the stop codon. This will include the full 3' UTR, but may decrease the efficiency of HDR.

Append 5'-TCAGTTAGCCTCCCCCATCTC-3' to the 3'-end of the downstream homology arm for C-terminal tagging using the insert from pMGS54. This sequence contains the template strand of the HygR coding sequence.

Phosphorothioate moieties are added to the 5' end of both upstream and downstream primers. PCR with the above primers using pMGS54 as template produces an amplicon of 1791 plus the length of the homology arm.

**Designing Primers for N-terminal tagging of the protein**

The 5' UTR can be important in the regulation of gene expression, therefore any changes in the sequence should be avoided. Select sgRNAs that cut inside the gene after or very proximal to the start codon.

**Upstream homology arm (coding strand primer design)**

*Cut site downstream of the start codon*

Start the homology arm 50 bases upstream of the cut site and end at the cut site. If the cut site is in the middle of a codon, ensure the HygR gene is in frame with the protein by adding extra nucleotides.

Append the sequence given below so that the primer anneals to the HygR-P2A-AID cassette in pMGS58.

*Cut site upstream of the start codon*

Start the upstream homology arm 50 bases upstream of the cut site and end at the nucleotide immediately preceding the start codon. Make necessary changes in the PAM sequence or the targeting sequence to avoid repeated cutting by the repaired insertion. However, avoid these changes if possible to mitigate the risk of altering protein expression.

Append the sequence given below so that the primer anneals to the HygR-P2A-AID cassette in pMGS58.

Append 5'-ATGAAAAAGCCTGAACTCACCG-3' to the 3'-end of the upstream homology arm for N-terminal tagging using the insert from pMGS58. This sequence contains the beginning of the HygR coding sequence.

**Downstream homology arm (template strand primer design)**

### *Cut site downstream of the start codon*

The downstream homology arm begins 50 bases downstream of the cut site and the last base is the nucleotide immediately downstream of the last start codon base. If necessary, change the wobble nucleotides of the codons before the cut site.

Append the sequence given below so that the primer anneals to the HygR-P2A-AID casette in pMGS58.

*Critical: confirm that the HygR-P2A-AID cassette is in frame with the protein after repair by in silico PCR using the SnapGene or any other program.*

### *Cut site upstream of the start codon*

The downstream homology arm begins 50 bases downstream of the cut site and the last base is the nucleotide immediately downstream of the last start codon base.

To functionally separate the protein of interest from AID, a linker of 9 amino acids is added at the C-terminus of the AID in the pMGS58 plasmid. The provided primer (below) amplifies both the linker and the AID. If any other template is used for generating the tag, be sure to add linker amino acid sequence.

*Critical: confirm that the HygR-P2A-AID cassette is in frame with the protein after repair by in silico PCR using the SnapGene or any other program.*

Append 5'-CCCACCTGAACCTCCAGATC-3' to the 3'-end of the downstream homology arm for N-terminal tagging using the insert from pMGS58. This sequence is complementary to the coding sequence of a flexible linker sequence following the end of the AID coding sequence in the plasmid. The PCR with the above primers using pMGS58 as template produces an amplicon of 1815 base pairs plus the length of the homology arm.

Add phosphorothioate moieties to the first two 5' nucleotides of both upstream and downstream primers.

**PCR amplification of Homology Directed Repair construct**

2-42    Synthesize the designed primers from Integrated DNA Technologies (IDT), adding phosphorothioate as a modification to the primer in the details.

2-43    Amplify the HDR template using the primers and Platinum Taq DNA Polymerase High Fidelity. We perform several 50 µl reactions (typically 4 to 8), and gel purify the PCR products.

**PCR mix:**

| | |
|---|---|
| Plasmid DNA | 50.0 ng |
| 10 µM Primer F | 1.0 µl |
| 10 µM Primer R | 1.0 µl |
| 10XHigh Fidelity buffer | 5.0 µl |
| DMSO | 0.5 µl |
| 10mM DNTP | 2.0 µl |
| $MgSO_4$ | 2.0 µl |
| Platinum Taq DNA Polymerase High Fidelity | 0.5 µl |
| Water | to 50 µl |

**Use PCR condition as follows:**

Initial denaturation
95°C 5 min

30 cycles of
95°C 30sec
60°C 30 sec
68°C 1 min

Final extension
68°C 10 min
4°C hold ∞

*Removing any remaining primer from the PCR products is important as primers may interfere with homologous recombination. The primers are very long, so conventional PCR clean-up kits will not remove primers efficiently and reduce the HDR efficiency by binding to the cut site. Always gel purify the PCR products.*

**Agarose gel purification**

2-44    Run the PCR products on an agarose gel and cut out the repair construct band.

2-45   Combine all the gel slices into one tube and purify the DNA using the Qiagen gel purification kit similar to the section 2-2 – 2-12.

2-46   Elute DNA using 50 to 100 µl nuclease-free water and quantify using a NanoDrop and store at -20°C. DNA is stable in -20°C for several months.

## AID-tagging the gene of interest

The three main steps in the tagging of a gene with AID include: 1) cotransfection of a gene-specific sgRNA and HDR template into ARF-TIR1 progenitor cells; 2) selecting tagged clones with hygromycin B; and 3) clonal expansion and confirmation of tagging. A general outline of these steps are illustrated in Figure 8.

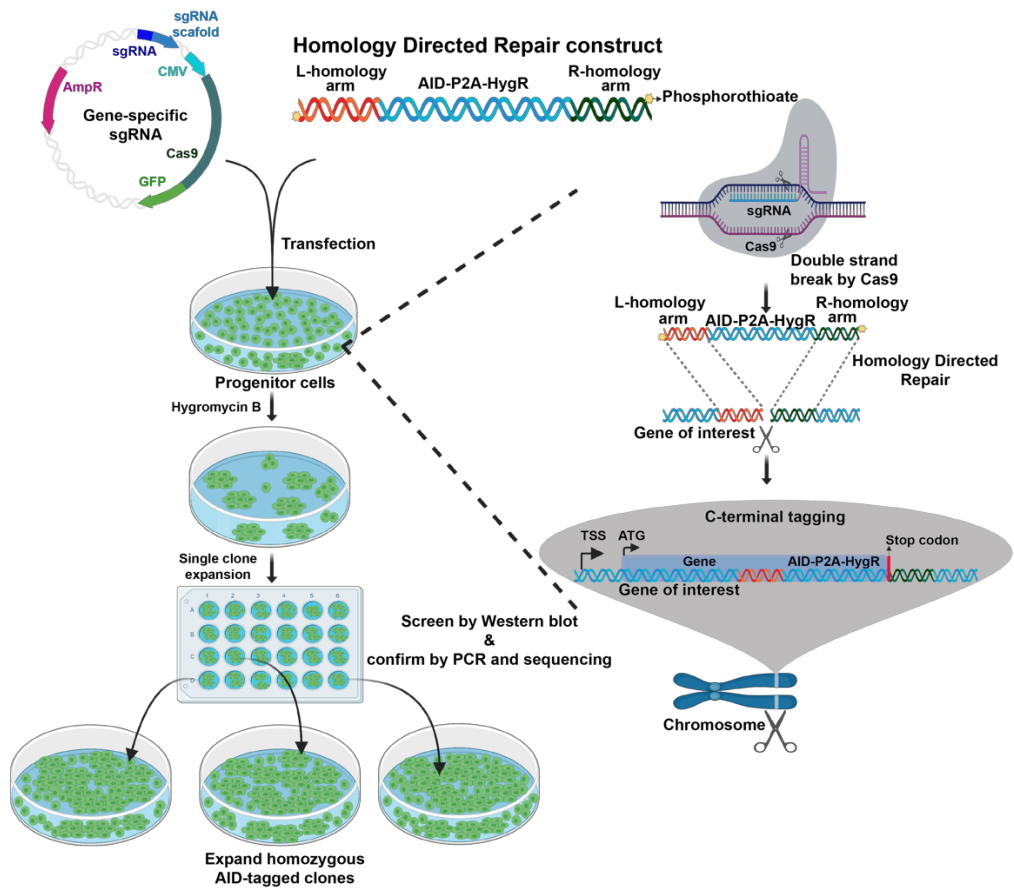

Figure 8. **HDR-mediated integration of AID-tag at the 3' end of the gene of interest using CRISPR-Cas9.** HEK293T-eGFP-ARF-P2A-TIR1 cells are cotransfected with an sgRNA and a PCR amplified repair construct. The cells are selected with hygromycin B and the clones are screened for integration using Western blotting, PCR, and sequencing.

**Transfection**

2-47   Grow the ARF-AID progenitor cells and split into six-well plates to ~30% confluency.

2-48   Cotransfect 1 μg of gene-specific sgRNA plasmid and 400 ng of double-stranded homology repair template PCR product to the cells as described in sections 1-1 – 1-9. Use 5 – 7.5 μl Lipofectamine reagent for transfection. The parental pX458 cotransfected with HDR PCR product is used as a transfection control and we recommend keeping a untransfected control cell well.

> *Transfect multiple (at least 4 wells) wells to get enough colonies.*

2-49   Replace with new media 24 hours after transfection.

2-50   After forty-eight hours, expand each well into a 10 cm plate.

> *Expanding cells into 10 cm plates spreads cells so they form isolated colonies.*

2-51   Seventy-two hours after transfection, add 100 μg/ml final concentration of hygromycin B to the cells and keep in the selection media until all cells in the untransfected control condition die. Antibiotic selection concentration varies between cell types. We recommend plotting a titration vs. cell viability curve to determine the lowest concentration at which nearly all cells die within 7 -12 days.

> *If the cell number is low after 72 hours, start hygromycin B selection on the 5th day of transfection. This increases the chance of getting positive colonies. The addition of conditioned media (8:2 ratio of new media with filtered cultured media from the same type of cell) increases the survival of the clones.*

2-52   After all the cells in the control plate are dead, replace the media with regular conditioned media in all the transfected cells and control plates.

2-53   Colonies will appear in the plates after two or three weeks.

2-54   Pick colonies similar to the section 1-15 – 1-22 and screen for tagged clones by Western blotting.

> *We use gene specific antibodies for detecting tagged proteins. The successful tagging results in an increase in protein size of 24 KD. Heterozygous integration will contain bands reflecting both the native and tagged proteins. The non-homologous end repair may produce read through into the 3' UTR or introduce stop codons before normal stop*

35 · ARF-AID system · arXiv

> codon of the gene. This is observed as an unexpected size shift that is not ~24KDa. Confirm the integration and reading frame of the integrated AID-P2A-HygR or HygR-P2A-AID by PCR and sequencing.

2-55   Select colonies from the Western blotting experiment and perform genomic DNA PCR using the same set of primers used for testing sgRNA efficiency according to sections 2-31 – 2-34, with the following change: PCR extension time should be increased to 4 minutes to amplify the insert.

2-56   Electrophorese whole PCR product on 1% agarose gel, excise the band using a clean scalpel and use the Qiagen gel purification kit to isolate genomic DNA (follow section 2-2 – 2-12). Sequence the purified PCR product with the forward primer to confirm the reading frame.

Successful integration of the AID-P2A-HygR or HygR-P2A-AID results in an addition of approximately 1785 bp to the PCR product. Heterozygous clones will have two bands, one with a length of genomic region between forward and reverse primer and the second with a length of genomic region between primers plus the 1785 bp. The homozygous integration will result in one band with a length of genomic region between primers plus the 1785 bp. Sequence the integrated DNA using the same forward primer, which will confirm successful integration.

## ALTERNATE PROTOCOL 1

**Establishment of the AID-ARF clamp system**

***Materials:***

pMGS59 (AID-ARF-P2A-Hygromycin), Addgene# 138174
pMK232 (CMV-OsTIR1-PURO), Addgene# 72834
pMGS7 (AAVS1sgRNA) Addgene #126582
ZNF143 C-terminal targeting sgRNA 5'-GAGGATTAATCATCCAACCC-3'

**ZNF143 C-terminal Homology Directed Repair Construct PCR Primers**
Forward 5'-
A*A*GAAGCCATCAGAATAGCGTCTAGAATCCAACAAGGAGAAACGCCAGGGCTTGACGACGGTGGATCTGGAGGTTCAGGTGGCAGTGTCGAGCTGAATCT-3'

Reverse 5'-
A*A*GACTCCTTCTGCTTTATTGCTCCATTGTTCTGAGGATTAATCATCCAATCAGTTAGCCTCCCCCATCTC-3'

This method uses the canonical TIR1 progenitor cells without the ARF protein. If the TIR1 expressing progenitor cells are available, directly tag the protein of interest with the AID-ARF clamp. Generate a TIR1 expressing progenitor cell using the TIR1 plasmid developed by the Kanemaki lab (Natsume et al., 2016) (Addgene# 72834) and the sgRNA that targets the AAVS1 locus (Addgene# 126582). Follow the steps 1-1 to 1-44 to make progenitor cells with the exception that this construct does not express eGFP-ARF.

    Tag at the C-terminus of the protein of interest with the AID-ARF clamp using plasmid #138174 from Addgene.

    We tagged ZNF143 at the C-terminus with the AID-ARF fusion protein (Figure 4) using the sgRNA and the donor primers given in the reagent list.

To tag protein of interest with ARF-AID clamp in the TIR1 progenitor cells, follow the steps 2-1 to 2-56. The only differences are the progenitor cells (TIR1 as opposed to ARF/TIR1) and the HDR template. Use the AID-ARF-P2A-Hygro plasmid (Addgene # 138174) to generate the HDR template.

For N-terminus tagging, the order of the AID and ARF fusion and linker properties should be empirically determined.

BASIC PROTOCOL 3

**TESTING AUXIN-MEDIATED DEGRADATION OF THE AID-TAGGED PROTEIN**

*Materials:*

Auxin (3-indole-acetic acid Sodium salt) (Abcam ab146403)

First, check the level of expression of the tagged clones compared to the progenitor cells by Western blotting. Run a serial dilution of the progenitor cell lysate along with all positive clones. Quantify the level of expression of the tagged proteins compared with that of untagged progenitor cells.

Second, test for rapid degradation of AID-tagged protein by adding 500 µM auxin directly into the culture media. Perform a treatment time course to determine the kinetics of degradation of the AID-tagged protein.

3-1    Make 50 mM auxin in water, aliquot and store at -20°C. The auxin is stable at -20°C for several months. Use a fresh aliquot each time and do not refreeze.

3-2    Seed 6-well plates with AID-tagged cells to achieve ~75% confluent cells the following day (plate approximately 7-8x10$^5$ HEK293T cells).

3-3    Add a final concentration of 500 µM auxin dropwise to the media all over the plate and mix by moving the plate forward and backward and sideways. Do not swirl the plate.

3-4    Collect cells at regular intervals starting from no auxin treatment. Initially, we collect at 15 minutes, 30 minutes, 1, 2, 3, and 4 hours. Remove one ml media from each well and collect cells by pipetting up and down in the remaining media. Put the cells on ice.

3-5    Centrifuge cells immediately after collection using a fixed angle rotor table top centrifuge at 6000 x g for 2 minutes at 4°C and carefully remove media by using a pipette.

> *Keeping cells on ice and centrifugation at 4°C significantly reduce further degradation of the tagged protein.*

3-6    Add 200 µl 2x SDS sample buffer directly into the pellet and pipette up and down several times. The lysate becomes highly viscous.

> *Alternatively, directly add the 2x SDS sample buffer into the plate after washing with PBS and lyse the cells. Collect the lysate with a pipette.*

3-7    Heat denature protein at 95°C for 5 minutes, vortex for 20 seconds and denature again for another 5 minutes. Store the lysate at -20°C.

3-8    Check the auxin-induced degradation of the tagged protein (Figure 3C-E, 4B&C). Serially dilute the untreated lysate to ensure that the query bands of the Western are within the linear range of the assay. Load the serial dilution of the untreated AID-tagged control lysate and include the treated cell lysate. Continue with Western blotting using antibodies directed against the AID-tagged protein. The degradation of the AID-tagged proteins starts immediately after adding auxin. Determine the rate of degradation by plotting the measured intensity of the AID-tagged protein bands using densitometry and fit the data using nonlinear regression and a one-phase decay equation.

Although the AID system works in many cell types and organisms, each cell type and organism is unique and the co-factors of the ubiquitin system may be differentially active.

## COMMENTARY

**TROUBLESHOOTING:**

**LACK OF POSITIVE COLONIES OR TOO MANY COLONIES DURING TAGGING**

If there is difficulty in getting positive colonies, check the efficiency of the sgRNA and try a different sgRNA if the efficiency is very low. Another potential problem is the disruption of the protein function by tagging with the AID. This is true for any essential genes that require both alleles for cell survival or if tagging makes a dominant negative mutant. The rate of homozygous integration is about 10% of the heterozygous integration in HEK293T cell, which has hyper diploid chromosome numbers.

If there are too many colonies and it is difficult to pick individual colonies, then split cells and plate around 100 to 200 cells per 10 cm plate. Depending on the cell type and cell survival after splitting, change the number of seeded cells. Grow cells with conditioned media to help individual cells to form colonies. Approximately 50 colonies in a 10 cm plate is optimal.

**TESTING THE FUNCTIONALITY OF THE TAGGED PROTEINS**

Absence of any tagged colonies may indicate that the tagged protein is not functional. If you are not able to generate homozygous clones after screening several clones (close to 100 heterozygous clones), then attempt to tag the protein at the other terminus. We recommend tagging a gene such as ZNF143 as a positive control, since ZNF143 is ubiquitously expressed and we previously optimized these sgRNAs and confirmed that C-terminally tagged ZNF143 is functional (Sathyan et al., 2019). To test the functionality of the tagged proteins, initially look at whether the protein localizes to the same compartment as the untagged proteins using immunofluorescence or cell fractionation. The same localization may indicate the protein is functional. For transcription factors, the localization to the same genomic loci is an indication of the functionality of a transcription factor binding and quantitative ChIP-seq can be used to determine if degradation results in genome-wide unidirectional decreases in binding (Guertin et al., 2018). Check the proximity of TF binding and the expression of the regulated genes after treatment with auxin. Depending upon the function of the protein, query the appropriate molecular phenotypes to confirm auxin-induced deficiencies.

**TIME CONSIDERATIONS:**

There are two components in the canonical AID, ARF-AID, and AID-ARF clamp systems. The generation of a progenitor cell and tagging the protein of interest with the degron. A general outline of the timeline to complete each step for HEK293T cells is given below, which may vary between cell lines used.

ARF-TIR1 or TIR1 progenitor line - 6-8 weeks
Design, clone, and test sgRNAs - 4 weeks.
Design and order the homology-directed repair constructs - 1 week
Tag the gene of interest with AID-tag - 6-8 weeks.

We recommend using established progenitor cells if available to reduce time to tag your protein of interest. Simultaneously developing progenitor cells, testing sgRNAs, and making homology-directed repair constructs significantly reduces the total amount of time to establish degron tagged cell lines.

**INTERNET RESOURCES**

sgRNA design tools:
| | |
|---|---|
| Benchling | https://www.benchling.com |
| CHOPCHOP | https://chopchop.cbu.uib.no |
| E-CRISP | http://www.e-crisp.org/E-CRISP/ |
| CRISPOR | http://crispor.tefor.net |

Primer design tools:
| | |
|---|---|
| Primer3 | http://bioinfo.ut.ee/primer3-0.4.0/ |
| IDT | https://www.idtdna.com/pages/tools/primerquest |
| Benchling | https://www.benchling.com |

**ACKNOWLEDGEMENTS**

We thank Arun Brendan Dutta, Anna Cetnarowska, Dr. Erin Catherine Moran and Dr. Piotr Przanowski for discussion and comments. Masato Kanemaki, Osaka University, provided the anti-TIR1 antibody, and Daniel Foltz, Northwestern University, provided the anti-GFP antibody. Illustrations were made using Biorender (https://biorender.com). This work was funded by the National Institute of Health GM128635 to MJG and MSTP training grant GM007267 to TGS.